\documentclass[aps,amsmath,twocolumn]{revtex4}
\usepackage{graphicx}
\usepackage{bm}
\begin{document}
\flushbottom

\title{Oxygen clamps in gold nanowires}

\author{
Frederico D. Novaes$^1$, Ant\^onio J. R. da Silva$^1$, E. Z. da Silva$^2$,
and A. Fazzio$^1$}

\affiliation{
$^1$Instituto de F\'\i sica, Universidade de S\~ao Paulo, CP 66318,
05315-970, S\~ao Paulo, SP, Brazil\\
$^2$Instituto de F\'\i sica ``Gleb Wataghin'', UNICAMP, CP 6165, 13083-970,
Campinas, SP, Brazil
}

\date{\today}

\begin{abstract}
We investigate how the insertion of an oxygen atom in an atomically thin gold nanowire can affect its rupture.
We find, using {\it ab initio} total energy density functional theory calculations, that O atoms when inserted in gold nanowires form
not only stable but also very strong bonds, in such a way that they can extract atoms from a stable tip, serving
in this way as a clamp that could be used to pull a string of gold atoms.
\end{abstract}


\maketitle

Gold nanowires have been at the center of many studies for the past 10 to 15 years\cite{landman90,muller92,agrait93,rubio96,landman96,ohnishi,yanson,kondo99,hakkinen,rubio01,edison01,ugarte01,takai,legoas02,unti,bahn,agrait03,novaes,skorod03,italia,nosso-prb,barnett,surfsci}. One of the reasons being that they can form chains
that are one atom wide, and a few atoms in length. These chains can be formed in a variety of ways; via a Scanning
Tunneling Microscope (STM) tip\cite{agrait93,rubio96,ohnishi,yanson} that contacts a surface and is then retrieved; from
a small junction that is mechanically broken\cite{muller92}; or through the bombardment of electrons that drill holes in a thin
gold foil\cite{kondo99,ugarte01}.
These works are motivated first for the fundamental interest to understand the properties of the ultimate wires, as well as the more applied question of how to use nanowires as electric contacts in nanoeletronic devices.
Ultimately, as Feynman pointed out in his seminal speech delivered at the APS 1959 Meeting: ``When we get to the very,
very small world - say circuits of seven atoms - we have a lot of new things that would happen that represent completely new opportunities for design.''

One observed fact is that atomically thin gold nanowires do not usually have more than three to four suspended
atoms connecting the leads.
In a previous study of the dynamical evolution of a gold nanowire, we investigated through tight-binding molecular dynamics (TB-MD) simulation the formation of such an atomically thin string of atoms all the way up to its rupture\cite{edison01}. We have found that gold atoms were inserted into the wire from the bulk until the tip connecting the suspended Au atoms attained a stable configuration. When the force required to extract another atom from the bulk was larger than that necessary to cause the rupture of the wire, it broke.

Moreover, if we
start from this final wire configuration (stable tip plus suspended gold atoms), but for a length smaller than the one required to break the
wire, and insert impurities such as H, C, C-C, S, N, C-H and C-H$_2$ in between two Au atoms in the wire, we observe that the
Au-X-Au (X being any one of the impurities) bond seems to be stronger than an Au-Au bond, since the wire's rupture always happened
at an Au-Au bond\cite{novaes}. The stable tips, however, did not change and as a consequence the wire's number of atoms
did not change. As we report below the same is not true for oxygen. It seems that the Au-Au-O-Au-Au bond becomes so strong that upon pulling
more gold atoms get extracted from the already stable tip, increasing in this way the wire's length. Therefore, it might be possible
to judiciously use  gold-oxygen bonds as a clamp to pull longer strings of suspended atoms, longer than in the pure gold case. Ideas
along these lines have been proposed for hydrogen \cite{barnett} and sulfur \cite{kruger03}. These studies are what is being called nanowire chemistry, a new branch of nanocatalysis.

\begin{figure}[ht]
\includegraphics[angle=0,width=7.cm]{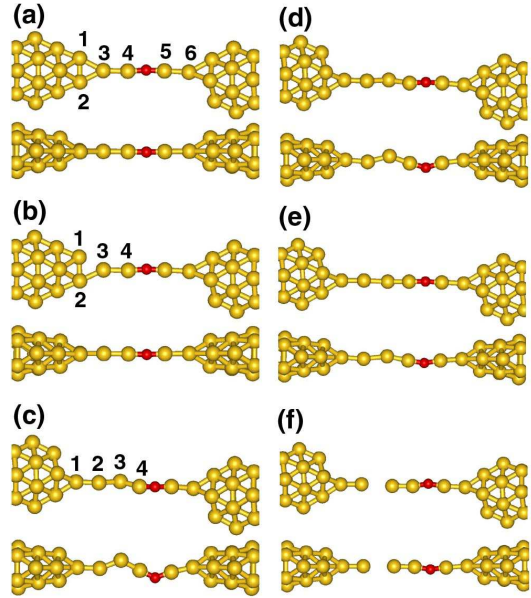}
\caption{(Colour online) Evolution of a gold (large spheres) nanowire with an incorporated oxygen (small
sphere) atom. Six representative configurations, (a)-(f), are presented. For each configuration two distinct views
of the wire are presented. The interatomic distances between labeled atoms are displayed in Table 1.}
\end{figure}

All our results are based on {\it ab initio} total energy density functional theory\cite{dft} calculations\cite{method}.
We have here employed the same procedure as in the previous previous investigation of the effect of impurities on the breaking of gold nanowires\cite{novaes,surfsci}.
The oxygen atom was
positioned in place of the Au atom at the center of the chain, forming in this way two Au-O bonds. Therefore, the wire now had
69 Au atoms and one O atom. After this insertion, all the atoms were again allowed to relax. We then pulled the wire by increasing
the supercell length, always allowing the whole system to relax between pulls. This protocol was repeated until we observed
the full rupture of the wire. It should be stressed that this procedure uses a realistic structure for the nanowire, in particular
where the impurity atom is placed, considers the wire under stress, and allows the system to break either at an Au-Au bond,
as in a clean wire, or at an Au-O-Au bond.

\begin{figure}[ht]
\includegraphics[angle=0,width=8. cm]{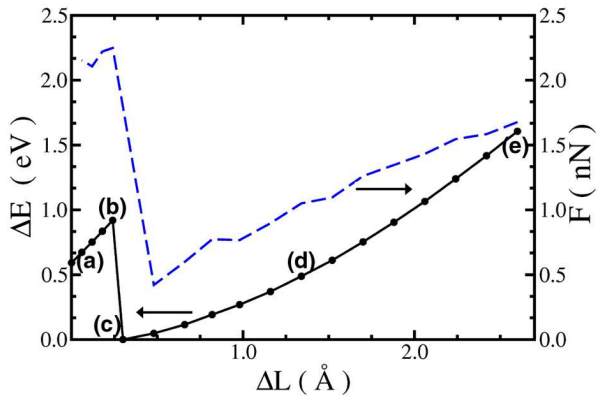}
\caption{
Total energy (lower curve; left axis) and calculated pulling force acting on the nanowire
(upper curve; right axis) along the stages of the simulation. The total energy values
are displayed relative to the lowest energy structure, shown in Fig. 1(c). The results
are presented as a function of the displacement $\Delta$L of the wire. The configurations
presented in Fig. 1 are marked accordingly in the graph.}
\end{figure}

The evolution of the gold nanowire with the incorporated oxygen atom may be seen in
Figure 1, where we present six representative structures. As can be seen from Fig. 1(a),
the atomically thin neck had originally four suspended gold atoms (labeled from 3 to 6).
The oxygen atom was inserted in the middle of the neck, between atoms 4 and 5. The interatomic
distances are presented in Table I. From the distances between the Au atoms
1 and 3 (3.06 \AA) and between the atoms 2 and 3 (2.97 \AA) in the configuration shown in Fig. 1(a),
and based on simulations\cite{novaes,nosso-prb} where
we noticed that the wire tends to break when there is an Au-Au distance close to 3.0-3.1 \AA,
we could expect that the wire would break. 
Instead of breaking, however, upon further pulling of the wire we have observed the rearengement of the tip,
with the removal of two atoms (atoms labeled 2 and 3). It can be seen from Fig. 1(b) the breaking of the bond between atoms 1 and
3, and subsequently (Fig. 1(c)) the gold atom 2 being also pulled from the tip into the
neck. This occured concomitantly with the restructuring of the tip, with the displacement of atom 1,
as can be observed in Fig. 1(c). One should note that oxygen caused all this sequence of events
even though the original geometry of the tip has been shown\cite{edison01,nosso-prb} to be quite stable,
a fact that happened only for the O atom among all the impurities that we have studied\cite{novaes,italia}
so far. This means that the oxygen impurity modifies and strengthens the neck significantly, which
is most likely related to strong Au-O bonds that are formed in low dimensional Au structures\cite{bahn}.

\begin{figure}
\includegraphics[angle=0,width=8.cm]{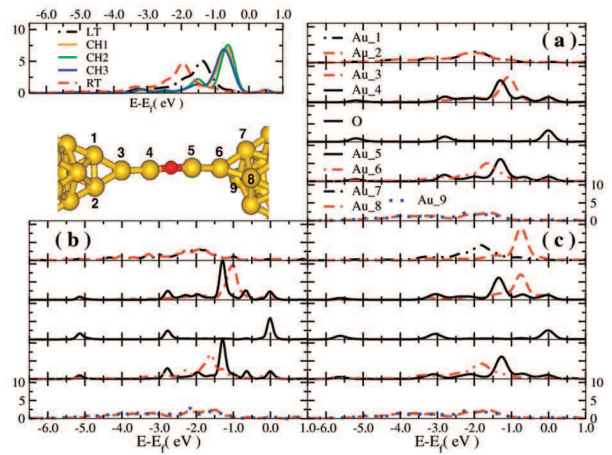}
\caption{(Colour online) Projected DOS (PDOS - arbitrary units) for the nine labeled gold atoms shown in the left middle panel,
plus the oxygen. In the top left panel we present the PDOS for the pure gold nanowire shown in Fig. 4(d). LT and RT
stand for left and right tip connecting atoms, respectively, whereas CH1, CH2 and CH3 correspond
to the first, second and third chain atoms, from left to wright, respectively. The results shown in figures (a) through (c) correspond to the structures presented in Fig. 1(a), Fig. 1(b), and Fig. 1(e), respectively.
}
\end{figure}

\begin{table}
\caption{Interatomic distances (in \AA) between Au atoms labeled from 1 to 6 in Fig. 1, for selected
snapshots (from Fig. 1(a) through Fig. 1(f)). The boldface distance for structure (f) indicates the bond that breaks.}
\label{tab1}
\begin{tabular}{ccccccc}
Structure & 1-2 & 1-3 & 2-3 & 3-4 & 4-5 & 5-6 \\
\hline
(a) & 2.71 & 3.06 & 2.97 & 2.82 & 4.20 & 2.82\\
(b) & 2.71 & 3.17 & 2.99 & 2.84 & 4.21 & 2.84\\
(c) & 2.65 & - & 2.66 & 2.59 & 3.70 & 2.61\\
(d) & 2.71 & - & 2.73 & 2.63 & 3.89 & 2.66\\
(e) & 2.79 & - & 2.98 & 2.69 & 4.07 & 2.73\\
(f) & 2.63 & - & {\bf 3.86} & 2.58 & 3.95 & 2.66
\end{tabular}
\end{table}

As a result of the insertion of more Au atoms in the neck, there is an overall stress release in the system. This can be
clearly seen in Fig. 2 where we present both the total energy as well as the calculated pulling force acting on the nanowire.
This is the expected behavior at this stage of nanowire stretching \cite{rubio01}, namely, an elastic stage with linear
increase of the force, followed by a sudden release of the tension related to the tip reconstruction. Fig. 1 and 2 show
exactly that. From a structural point of view, what happened after the tip reconstruction was the
formation of a zig-zag structure in the neck, as can be seen in Fig. 1(c).  This is a consequence of the smaller stress that
is being applied to the neck region as a result of the insertion of more atoms in the neck.
During the elastic stage following the stress release, there is a ''linearization'' of the neck, that continues all
the way up to the rupture point, that occurs between atoms 2 and 3. This happens for a pulling force of
approximately 1.7 nN, which is in good agreement with the experimental value of 1.5 $\pm$ 0.3 nN measured for a pure
gold nanowire\cite{rubio01}. This value for the rupture force is what one would expect for the breaking of an Au-Au bond.
This can be compared to the value of 2.3 nN of the force that causes the tip reconstruction,
indicating that the whole neck, including oxygen second nearest neighbor pure Au-Au bonds, are strengthened by the
presence of the O atom.

\begin{figure}
\includegraphics[angle=0,width=8.cm]{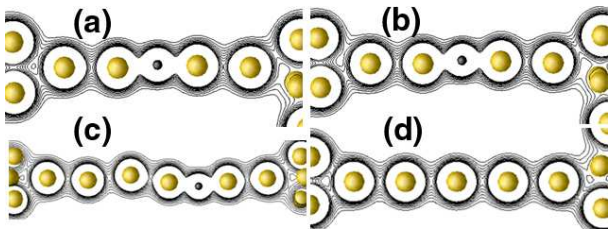}
\caption{(Colour online) 
Contour plots for the total electronic charge density, in planes that pass through the neck atoms.
The results shown in figures (a) through (c) correspond to the structures presented in Fig. 1(a), Fig. 1(b), and
Fig. 1(e), respectively. In (d) we present a similar plot for a pure gold nanowire prior to its rupture.}
\end{figure}

In order to gain insight into the electronic structure of the Au nanowires with the O impurity, and how it evolves
as the wire is pulled, we present in Fig. 3 the projected density of states (PDOS).  The PDOS is shown for all gold
atoms labeled from 1 to 9 in the left middle panel of Fig. 3, which encompasses all the neck atoms as well as some of the tip
atoms. The PDOS for the oxygen atom is also shown. In Fig. 3 (top left panel) we also present, for comparison, the PDOS for
a pure gold nanowire (see Fig. 4 for the atomic structure). In Fig. 3(a) we show the PDOS for the configuration
shown in Fig. 1(a). For this configuration the tip atoms (atoms 1,2,7,8, and 9) have an electronic structure which
is similar to bulk gold, as we had already observed before in pure nanowires\cite{nosso-prb}. Again, similarly to the
pure gold nanowire, for the neck atoms (atoms 3-6) that have a smaller number of neighbors, we observe:
(i) that their PDOS become narrower; (ii) are characterized by sharper peaks; and (iii) present an overall shift towards
higher energies. The fact that the Au PDOS has an overall shift towards lower energies as the number of
nearest neighbors increases can be clearly seen when one compares the results for atoms 3 and 6, that connect the
tip to the neck. Atom 3 has three neighbors whereas atom 6 has four neighbors. This is reflected in a small shift of the
atom 6 PDOS towards lower energies when compared to the atom 3 PDOS. This can be also seen in
the pure gold nanowire PDOS. Finally, the oxygen atom PDOS has a clear
peak at the Fermi energy. The O two gold neighbor atoms also have peaks at the Fermi energy, a result of the chemical
bonding and orbital hybridization. In fact, the PDOS of these two Au atoms (atoms 4 and 5) are very similar.
An important feature related to the wire's stability is the peak at approximately 5 eV below the Fermi level. From
charge density plots (not shown) it is possible to see that this peak is associated  with a $O_p-Au_d$ bonding orbital. This
orbital clearly involves the oxygen atom and its two Au nearest neighbors, however, it also has some weight from
the oxygen Au second nearest neighbors. These peaks are clearly absent in the pure Au PDOS, indicating that
besides the delocalized, metal-like bonding that stabilizes these Au wires, the local, covalent-like bonding between
the Au and O have a significant contribution in the strength increase that we observe.

In Fig. 3(b) we present the PDOS for the configuration shown in Fig. 1(b). As the wire is stretched,
there is a sharpening of the projected DOS of the neck atoms, in particular for atom 3, since it is evolving
from a three-atom coordination to a two-atom
coordination. There are no significant alterations in the PDOS for the tip atoms. In particular,
the PDOS for the tip atoms 7, 8, and 9 do not suffer any significant changes throughout the evolution
of the wire, since their environment does not get altered.
The PDOS for the configuration shown in Fig. 1(e) is presented in Fig. 3(c). Most of the features already
discussed are also present. The only noticeable changes when compared to Fig. 3(a) are the
significant modifications in the PDOS of atom 2. As this atom is now a neck atom, its PDOS is shifted
towards higher energies, and develops a sharper peak when compared to a more bulk like atom.

In Fig. 4 we show contour plots for the total electronic charge density, in planes that pass through the neck atoms.
Fig. 4(a) corresponds to the structure presented in Fig. 1(a). The charge density for the Au-O bonds
when compared to the value for the pure Au-Au bonds indicate that oxygen is in fact making
strong bonds with the gold atoms. This feature is preserved as the wire evolves, as can be seen in
Fig. 4(b), which corresponds to the structure presented in Fig. 1(b), and in Fig. 4(c), which corresponds to the
structure presented in Fig. 1(e). Moreover, there is also an increase of the charge density
between the Au-Au bonds closest to the oxygen atom (between atoms 3 and 4 and between atoms
5 and 6). Therefore, it seems that the oxygen atoms is affecting the wire in such a way that
the structure Au-Au-O-Au-Au becomes strong. For comparison, we show in Fig, 4(d) similar
contour plots for a pure gold nanowire prior to its rupture. From Fig. 4(a) and Fig. 4(b) we can see that
the bond between atoms 1 and 3 is becoming weaker, which is consistent with the restructuring
of the tip. Also, from Fig. 4(c) the weaker bond is between atoms 2 and 3, which is precisely where
the wire breaks.

From all these results we can rationalize some general features: (i)  even though there is a metallic, delocalized
character in the bonds at the nanowire, they also have a rather localized nature involving nearest neighbor atoms.
This is evident both from the fact that gold atoms with similar number of neighbors have similar PDOS and from the
observation that the impurity affects primarily the PDOS of its nearest neighbors. Also, there are clearly
Au-O-Au bonding orbitals that show up as low energy peaks in the PDOS; (ii) the rupture of the wire usually
happens at bonds involving atoms whose PDOS are displaced towards higher energy values,
closer to the Fermi energy, which is related to a weakening of these bonds. See, for example, the PDOS for
atoms 2 and 3 in Fig. 3(c). In particular, the PDOS for atom 3 is also closer to the Fermi energy in Fig. 3(a), and
the first bond that breaks at the tip reconstruction is precisely between atoms 3 and 1. However, the
PDOS for atom 1 is still delocalized because it has a higher coordination number; (iii) the oxygen atom has
an influence that extends towards its next-nearest neighbors. This is why the bond between atoms 3 and 4 does
not break. Instead, the neck bonds become so strong that the pulling force attains a value of 2.3 nN,
which is suficcient to disrupt the tip.

In conclusion, we have presented state-of-the-art calculations of a realistic Au atomic chain nanowire with an inserted
oxygen impurity atom. In particular, we have investigated how the O atom can affect the rupture of the wire.
We find that O atoms can form not only stable but also very strong bonds with the Au atoms, in such a way that they
can extract atoms from a stable tip. This indicates that in an oxygen reach atmosphere it may be possible to
pull a longer string of gold atoms\cite{jan}. If one considers that in most situations wires with more than, say, 4-5 atoms
have not been observed, the probable reason being that stable tips are quickly formed and
Au-Au bonds do not sustain forces greater than approximately 1.8 nN, this opens up the possiblity to study:
(i) if this local bond strength interpratation is valid; (ii) how stable can tips be; and (iii) the general stability
properties of Au nanowires. Finally, as the O atom has a clear peak at the Fermi energy, it will be interesting to
see how it affects the charge transport properties of the nanowire. Such an investigation is currently under way.

We thank J. M. van Ruitenbeek for communicating unpublished work.
We acknowledge support from FAPESP, Capes and CNPq.


\begin{references}
\bibitem{method}
We have used a GGA approximation\cite{pbe}; norm-conserving pseudopotentials \cite{pseudo};
periodic-boundary conditions; a supercell approximation with lateral separation of 20 $\text{\AA}$ between wires;
and the X-point for the Brillouin zone sampling. All calculations were performed using the SIESTA code\cite{siesta}, and we employ a DZP basis function with a confining energy shift of 0.07 eV\cite{basis} and a cutoff of 250 Ry for the grid integration. The procedure of elongation and atomic relaxation was the same as in \cite{novaes}. Even though
we report non-spin polarized calculations, we have performed tests including spin polarization for geometries close to the tip reconstruction point, and none of the main conclusions have changed.
\end{references}
\end{document}